\documentclass[pra,twocolumn]{revtex4-1}
\usepackage{xcolor}
\definecolor{maroon}{RGB}{150,0,0}
\definecolor{dblue}{RGB}{65,105,250}
\usepackage[colorlinks=true,linkcolor=black,citecolor=maroon,urlcolor=maroon]{hyperref}
\usepackage{graphicx,multirow,amsmath}

\newcommand{\ie}{{\it i.~e.~}}
\begin{document}
\title{On-demand quantum key distribution
using superconducting rings with a mesoscopic Josephson
junction}
\author{Chandan Kumar}
\email{chandankumar@iisermohali.ac.in}
\affiliation{Department of physical sciences, Indian
Institute of Science Education and Research (IISER), Mohali,
Sector 81 SAS Nagar, Manauli PO 140306, Punjab, India}
\author{Jaskaran Singh}
\email{jaskaransinghnirankari@iisermohali.ac.in}
\affiliation{Department of physical sciences, Indian
Institute of Science Education and Research (IISER), Mohali,
Sector 81 SAS Nagar, Manauli PO 140306, Punjab, India}
\author{Arvind}
\email{arvind@iisermohali.ac.in}
\affiliation{Department of physical sciences, Indian
Institute of Science Education and Research (IISER), Mohali,
Sector 81 SAS Nagar, Manauli PO 140306, Punjab, India}
\begin{abstract}
We present a quantum key distribution (QKD) protocol based
on long lived coherent states prepared on superconducting
rings with a mesoscopic Josephson junction (dc-SQUIDs). This
enables storage of the prepared states for long durations
before actually performing the key distribution.  Our
on-demand QKD protocol is closely related to the coherent
state based continuous variable quantum key distribution
protocol.  A detailed analysis of preparation, evolution and
different measurement schemes that are required to  be
implemented on dc-SQUIDs to carry out the QKD is provided.
We present two variants of the protocol, one requiring time
stamping of states and offering a higher key rate and the other
without time stamping and a lower key rate.  This is a step
towards having non-photon based QKD protocols which will be
eventually desirable as photon states cannot be stored for
long and therefore the key distribution has to be
implemented immediately after photon exchange has occurred.
Our protocol offers an innovative scheme to perform QKD and
can be realized using current experimental techniques.
\end{abstract}
\maketitle
\section{Introduction}
\label{sec:intro}
In the quantum world the joint measurement  of non-commuting
observables is impossible and if attempted, leads to the
introduction of disturbances in the measured outcomes for
both. This fundamental feature of quantum physics was
exploited by Bennett and Brassard to invent a  secure key
distribution protocol for cryptography~\cite{BB84, B92}.
Unlike classical key distribution schemes where the security
is based on `a hard to solve mathematical problem', security
of quantum key distribution (QKD) protocols is based on
laws of nature which cannot be
violated~\cite{Crypt_review,Shor_Pres_2000, Sec_BB84}.
Subsequently, other key quantum features, like
entanglement~\cite{Crypt_review},
Bell-nonlocality~\cite{E91,Acin_sec}, no-cloning
theorem~\cite{Cloning_sec,cv_qkd_ralph} and monogamy of
quantum correlations~\cite{Sec_monogamy,jask_mono} have also
been employed to show unconditional
security~\cite{Sec_qkd,sec_cv_ralph,coh_qkd} of discrete and
continuous variable QKD protocols.

The experimental realizations of QKD protocols have
primarily been on light and optical
devices~\cite{expt_cv_qkd,expt_qkd1,expt_qkd2,expt_qkd3,expt_qkd4}.
While the optical setup has many advantages, it imposes a
serious constraint: Bob has to perform his chosen
measurement immediately once he receives the state because
it is not possible to store photonic states for long
durations. In this paper we develop a novel way to perform
QKD in which Bob is not under any compulsion to perform
measurements as soon as the states are exchanged.  Bob may
choose to store the states until the need for key
distribution arises, at which time he performs measurements,
exchanges classical information with Alice and generates the
key.  The proposed protocol is a continuous variable QKD
based on superconducting rings with a Josephson
junction~\cite{junction,squid_ring,squid}. This system, known
as dc-SQUIDs, allows for preparation of extremely long lived
continuous variable coherent states under no-dissipation
conditions~\cite{squid_hamil,squid_hamil_diss}. These states
can be stored as such indefinitely at equilibrium, which
ensures a dissipationless situation~\cite{Friedman2000}. 
For sufficiently low temperatures, a persistent current has
been shown to exist in these families of superconducting
rings with mesoscopic Josephson
junctions~\cite{dissipationless1,dissipationless2}

The protocol involves preparing arbitrary coherent
states of the dc-SQUID by Alice, which are then transported
to Bob.  Bob stores these states for as long as he wants,
during which time these states evolve under the system
Hamiltonian.  Bob undertakes measurements on these dc-SQUID
states when he wants to generate the key.  We find that the
states during the storage time undergo collapse and revival
phenomena due to the presence of a nonlinear term in the
Hamiltonian.  Given this, two distinct measurement schemes
are possible, either of which can implemented by Bob to
generate the key. In the first scheme, Bob measures at an
arbitrary time and in the second scheme, he measures at
specified time intervals to maximize the secure key rate.
The second measurement scheme requires time stamping, \ie
Alice and Bob have to share a clock and a precise time of
preparation of the coherent state needs to be marked on each
dc-SQUID.  We show that a secure key rate of $0.20$ bits and
$0.50$ bits can be achieved for the schemes, respectively.
Finally, we show that the protocol is secure against an
eavesdropper under the assumption of individual attacks. The
new feature of the protocol is the possibility of storing
the dc-SQUID, prepared in a coherent state, in Bob's lab for
long durations and the use of a non-photonic system for QKD.
The proposed protocol can be implemented using current
superconducting technologies. Since the QKD can be carried
out at any time after state exchange, we call our protocol
as on-demand QKD.

The paper is organized as follows: In Section~\ref{sec:back
a} we briefly introduce the concept of continuous variable
quantum key distribution (CV-QKD) with a focus on coherent
states, while in Section~\ref{sec:squid} we review basic
concepts of superconductivity as pertaining to
superconducting rings with a mesoscopic Josephson junction.
We describe the preparation and evolution of these states in
some details in Sections~\ref{sec:preparation}
and~\ref{sec:storage}.  In Section~\ref{sec:protocol} we
describe  our QKD protocol while in 
Section~\ref{sec:security} we prove its security.  
In Section~\ref{sec:conc}
we provide some concluding remarks.
%%%%%%%%%%%%%%%%%%%%%%%%%%%%%%%%%%%%%%%%%%%%%%
\section{Background}
\label{sec:back}
\subsection{Coherent state based CV-QKD}
\label{sec:back a}
In this section we provide a brief introduction to CV-QKD
with a focus on the coherent state protocol as detailed
in~\cite{coh_qkd}. We sketch out the protocol and briefly
analyze the security considerations.
 
Consider two parties Alice and Bob who wish to share a
secret key using a coherent state based QKD protocol. Alice
randomly draws two numbers $x_A$ and $p_A$ from two Gaussian
distributions with the same variance $V_AN_0$, where $N_0=1$
is the vacuum noise variance. She then prepares the coherent
state $|x_A+ip_A\rangle$ and transmits it to Bob through a
channel with Gaussian and white noise. Upon receiving the
state, Bob randomly chooses to perform a homodyne
measurement of either $X$ (position) or $P$ (momentum)
quadrature.

Through a classically authenticated channel Bob reveals his
choices of measurement quadratures to Alice, who then
rejects the random number not corresponding to the measured
quadrature. The procedure is repeated a large number of
times and the random number with Alice and the outcome with
Bob is kept as part of the raw key. In the end, Alice and
Bob can use the sliced reconciliation
protocol~\cite{Cloning_sec} to transform their raw key into
errorless bit strings from which a secure key can be
distilled by privacy amplification.

If we consider the presence of an eavesdropper, Eve in the channel,
a secret key can be distilled from the
protocol only if 
\begin{equation}
r_{\text{min}}=I(A:B)-I(A:E)>0, \label{eq:rmin}
\end{equation}
where $I(A:B)$ and $I(A:E)$ represent the mutual
information between Alice and Bob and Alice and Eve
respectively and $r_{\text{min}}$ is the minimum rate at which a
secure key can be distilled. Eqn.~(\ref{eq:rmin}) physically
implies that a secure key can be distilled only when the
information shared between Alice and Bob is strictly greater
than the information shared between Alice and Eve.

From a result in~\cite{shannon}, it can be shown that
if the signal and noise have Gaussian statistics, the
optimum information rate achievable
between Alice and Bob is
\begin{equation} I(A:B) =
\frac{1}{2}\log_2(1+\Sigma_B).  \label{eq:iab}
\end{equation}
where $\Sigma_B$ is the signal-to-noise ratio (SNR) as
measured by Bob. To evaluate the maximum information about
Alice's key gleaned by Eve, it is required to ascribe the
best physically possible strategy to her. If the
transmission line between Alice and Bob has transmittivity
$\eta$, Eve can then employ a strategy wherein she captures
a fraction $1-\eta$ of the beam and transmits the fraction
$\eta$ to Bob through her own lossless line. This way she
gains maximum amount of information.

A general result proven in~\cite{qkd_sec_nocloning}
demonstrates that if the added noise on Bob's side is $\chi
N_0$, the minimum added noise on Eve's side is
$\chi^{-1}N_0$, where $\chi = {(1-\eta)}/{\eta}$. The above
is a consequence of the no-cloning theorem and can be
applied to show security of the aforementioned protocol when
Eve performs individual attacks.

In the case of the coherent state protocol, the total
variance of the beam leaving Alice's site is $VN_0=V_AN_0+N_0$. The
security condition~(\ref{eq:rmin}) then reads as
\begin{equation} \begin{aligned}
r_{\text{min}} &=
\frac{1}{2}\log_2(1+\Sigma_B)-\frac{1}{2}\log_2(1+\Sigma_E)\\
&=\frac{1}{2}\log_2\left(\frac{V+\chi}{1+V\chi}\right),
\end{aligned} \label{eq:rsecure} 
\end{equation} 
where we have used
\begin{equation} 1+\Sigma_B =
\frac{V+\chi}{1+\chi}, \quad 1+\Sigma_E =
\frac{V+1/\chi}{1+1/\chi}. 
\label{eq:sigma} 
\end{equation}
From Eqn.~(\ref{eq:rsecure}) it is seen that
a secure key can be distilled if $\chi<1$. This further puts
bounds on the transmittivity $\eta>\frac{1}{2}$. 
In other words, as long as Alice-Bob channel transmission
efficiency is greater than 50\%, they can successfully 
carry out QKD.
%%%%%%%%%%%%%%%%%%%%%%%%%%%%%%%%
\subsection{Superconducting ring with a
junction (dc-SQUID)} 
\label{sec:squid}
\begin{figure}
\includegraphics[scale=1]{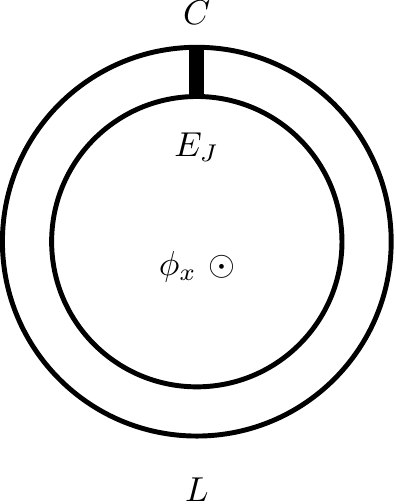}
\caption{A superconducting ring of inductance $L$ with a
mesoscopic Josephson junction with capacitance $C$. The
Josephson coupling constant is $E_J$ and $\phi_x$ is the
external flux.}
\label{fig:squid}
\end{figure}
%%%%%%%%%%%%%%%%%%%%%%%%%%%%%%%5
In this section we provide a brief background to
superconducting rings with a mesoscopic Josephson junction,
also termed as a dc-SQUID. We mainly focus on preparation of
coherent quantum
states~\cite{squid_hamil,squid_hamil_diss} and their
evolution on a dc-SQUID which is essential for our protocol.

Consider a superconducting ring of inductance $L$ with a
Josephson junction with capacitance $C$ and external
inductive coupling through an external flux $\phi_x$, 
as shown in Fig.~\ref{fig:squid}. The quantum
Hamiltonian for the junction can be written as,
\begin{equation} 
H = \frac{Q^2}{2 C}+\frac{(\Phi'-{\phi}_{x})^2}{2
L}+E_{J}(1-\cos\theta),
\label{eq:hamiltonian} 
\end{equation}
where we have taken $\hbar=k_B=c=1$. The quantity $Q$
is the charge
operator
across the junction
and  $\Phi'$ is the operator corresponding to
 the total flux through the ring. 
The quantity $E_J$ is the Josephson coupling
constant and
$\theta$ is the phase difference of the superconducting
wavefunction across the junction. The phase difference
$\theta$ is related to the total flux $\Phi'$ by $\theta=
2e\Phi'$. The quantum mechanical operators, $Q$ and $\Phi'$
form a canonically conjugate pair of variables with the
canonical commutation relationship
\begin{equation}
\left[\Phi', Q\right]=i.
\end{equation}
The commutation relation can be
rewritten in terms of voltage across the junction, $V'=Q/C$
as, 
\begin{equation} 
C\left[\Phi', V'\right]=i.
\label{eq:commutation} 
\end{equation}
Due to experimental considerations, in the remainder of this
paper we treat $V'$ and $\Phi'$ equivalent to the continuous
variable quantum mechanical quadrature operators rather than $Q$ and
$\Phi'$, as is evident from Eqn.~(\ref{eq:commutation}).
The corresponding uncertainty relationship for the $V'$ and
$\Phi'$ quadratures is,
\begin{equation}
\langle\left(\Delta \Phi'\right)^2\rangle\langle \left(\Delta
V'\right)^2 \rangle \geq \frac{1}{4C^2}.
\label{eq:uncertainty}
\end{equation}

The Hamiltonian for a dc-SQUID~(\ref{eq:hamiltonian}) is
found to be equivalent to that of a simple harmonic
oscillator with an additional coupling term proportional to
the Josephson coupling energy. The first term corresponds to
kinetic energy and the last two terms correspond to
potential energy. 

Expanding the last term in the Hamiltonian~(\ref{eq:hamiltonian})
 and retaining terms upto the fourth order in
$\theta$, we get,
\begin{eqnarray} 
H &=&
\frac{Q^2}{2 C}+\frac{C}{2}\omega^2
\Phi'^2 -\frac{\Phi' \phi_x}{L}-\frac{2}{3}E_{J}e^4{\Phi'}^4,
\nonumber \\
&&{\rm with}\quad 
\omega = \left(\frac{1}{CL} + \frac{4e^2E_J}{C}\right)^{\frac{1}{2}}. 
 \label{eq:omega_small}
\label{eq:hamiltonian_fourth_order} 
\end{eqnarray}
To simplify calculations we work with the dimensionless quadratures defined as,
\begin{equation}
\Phi = \sqrt{C\omega}\Phi' \quad \text{and} \quad V=\sqrt{\frac{C}{\omega}}V'
\end{equation}
The creation and
annihilation operators, $b$ and $b^\dagger$ can then be introduced
as:
\begin{equation} 
\Phi=\frac{1}{\sqrt{2}}(b+b^{\dagger}), \quad V=\frac{i}{\sqrt{2}}(b^{\dagger}-b).  \label{eq:creation} 
\end{equation}
In the rotating wave approximation and neglecting all the
terms without annihilation and creation operators,
Eqn.~(\ref{eq:hamiltonian_fourth_order}) can be re-written
as,
\begin{equation} H = \Omega b^{\dagger}b -
\mu (b+b^{\dagger})-\nu(b^{\dagger}b)^2, 
\label{eq:eff_hamiltonian} 
\end{equation}
where
\begin{equation} 
\nu = \frac{2E_Je^4}{3(\omega
C)^2}, \quad \mu = \frac{\phi_x}{L\sqrt{2 \omega C}}, \quad
\Omega= \omega-\nu.  \label{eq:nu_mu_omega} 
\end{equation}
The second term in the 
Hamiltonian~(\ref{eq:eff_hamiltonian}) 
depends on the external driving flux
$\phi_x$, which can be switched on for a short duration
$\tau_1$ when desired. The strength of the driving flux is
appropriately chosen such that $\mu\gg\Omega, \nu$ for the
duration when it is turned on.  We further impose the
experimentally achievable condition $\Omega\gg\nu$ and
for the remainder of the paper work in the regime
$\mu\gg\Omega\gg\nu$~\cite{squid_hamil,squid_hamil_diss}.

For the duration $\tau_1$ when the driving field is turned
on, only the second term of the 
Hamiltonian~(\ref{eq:eff_hamiltonian}) 
is relevant and it effectively generates a
phase space displacement of the ground state and its
corresponding unitary operator can be written as

\begin{equation} D(\tau_1) =
\text{e}^{i\mu(b + b^\dagger)\tau_1 }, 
\label{eq:displacement} 
\end{equation}
By a suitable choice of the external driving field $\phi_x$
and time durations $\tau_1$, we can modulate the value of
$\mu$ and prepare an arbitrary coherent state.  However, it
should be noted that $\mu$ is real and thus the
corresponding displacement operator~(\ref{eq:displacement})
will displace the ground state of the junction only along
the $V$ quadrature.

After the driving field is switched off, the
Hamiltonian consists of only the first and the third term.
From the condition $\Omega\gg\nu$, for a short duration
$\tau_2$ the first term dominates and the effective
Hamiltonian is responsible for generating a phase shift 
which corresponds to a phase space rotation with unitary
operator given as
\begin{equation} 
R(\tau_2)= \text{e}^{
-i\Omega b^{\dagger}b\tau_2}.  
\end{equation}
For a long storage time duration $\tau_3$, the
external field is switched off and the third term in the
Hamiltonian~(\ref{eq:eff_hamiltonian}) becomes significant
alongwith the first term. In the interaction picture where
the first terms gets absorbed, the unitary operator
corresponding to the Hamiltonian for the storage period is,
\begin{equation} 
S(\tau_3) = \text{e}^{
i\nu( b^\dagger b)^2 \tau_3}.  \label{eq:squeeze}
\end{equation}
The effect of the third non-linear term is to squeeze and
de-squeeze the coherent states of the Josephson junction in
the $V$ and $\Phi$ quadrature resulting in a collapse and
revival phenomenon. The system starts out in a coherent
state which then gets
 squeezed by the nonlinear term. At time
$\tau_3=\frac{ \pi}{\nu}$ the squeezing vanishes and the
resultant state is $|-\alpha\rangle$, 
as will be shown in Section~\ref{sec:storage}, which is just the initial
coherent state rotated by an angle $\pi$. As time elapses,
the state is again squeezed and at a time $\tau_3=\frac{
2\pi}{\nu} $ the squeezing again vanishes and the resultant
state is found to be $|\alpha\rangle$, which is exactly the
initial coherent state. This cycle of collapse and revival
continues indefinitely under no-dissipation conditions.  In
between, we also see superposition of two coherent states.
%%%%%%%%%%%%%%%%%%%%%%%%%%%%%%%%%%%%%
\subsection{Preparation of coherent states on a dc-SQUID} 
\label{sec:preparation}
Alice can prepare the chosen coherent state
$|\phi_A+iv_A\rangle$ on a dc-SQUID by first preparing it in
the ground state $|0\rangle$. This can be achieved by
allowing the SQUID to decohere to its ground state in the
low Josephson coupling limit~\cite{ground_state}.
By appropriately choosing the
magnitude of the driving field $\phi_x$, Alice applies the
displacement operator for a short duration $\tau_1$, given
by Eqn.~(\ref{eq:displacement}), on the ground state to get:
\begin{equation}
D\left(\frac{\phi_A}{\mu_1}\right)|0\rangle = |0+i\phi_A\rangle.
\label{eq:displ_pa}
\end{equation}
Since $\mu_1$ is a real quantity, the state is displaced
only along the $V$ quadrature. The driving field is then
switched off for a time $\tau_2$ such that $\Omega \tau_2 =
\pi/2$ which results in rotating the quadratures of the
coherent state by an angle $\pi/2$:
\begin{equation}
R\left(\frac{\pi}{2\Omega}\right)|0+i\phi_A\rangle=|\phi_A+0\rangle.
\label{eq:rotation}
\end{equation}
Since $\tau_2$ is quite small, the nonlinear term
in~(\ref{eq:eff_hamiltonian}) can be ignored as we have $\Omega\gg \nu$. The resultant state is then once
again displaced along the $V$ quadrature by appropriately
choosing the magnitude of the driving field for a short
duration $\tau_1$:
\begin{equation}
D\left(\frac{v_A}{\mu_2}\right)|\phi_A+0\rangle = |\phi_A + iv_A\rangle.
\label{eq:prepared_coherent}
\end{equation}

The entire procedure can be
summarized by an application of the operator $T(\phi_A,
v_A)$ on the ground state of the dc-SQUID as,

\begin{equation}
T(\phi_A, v_A) =
D\left(\frac{v_A}{\mu_2}\right)
R\left(\frac{\pi}{2\Omega}\right)D\left(\frac{\phi_A}{\mu_1}\right).
\label{eq:eff_creation_operator} 
\end{equation}
The operator $T(\phi_A,
v_A)$ depends on the structural properties of the junction,
the applied external driving field and the time durations
$\tau_1$ and $\tau_2$. By appropriately choosing these
values a dc-SQUID can be prepared in an arbitrary coherent
state as required. 
\begin{figure}
\centering
\includegraphics[scale=1]{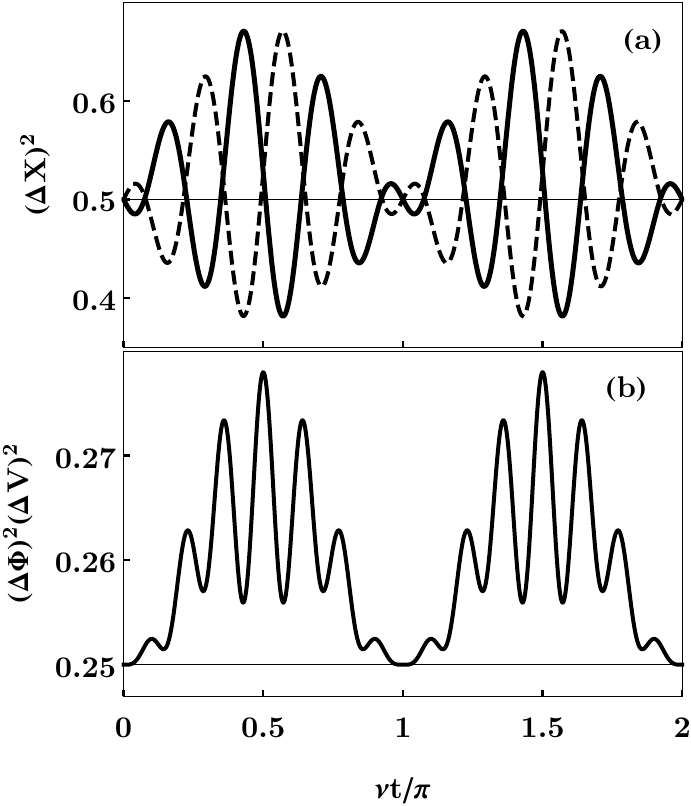}
\caption{($a$) Solid line represents the variance of $\Phi$
quadrature while the dashed line represents variance of $V$
quadrature. The condition for squeezing in quadrature $X$
is $\langle(\Delta X)^2\rangle<1/2$. It can be seen that
while one quadrature is squeezed, its conjugate is
de-squeezed.($b$) For state~(\ref{eq:resultant_state1}) the
product of the variance is always greater than 1/4 except
when the state revives to a coherent state. To clearly show
squeezing and desqueezing of quadratures, we have taken
$\Omega= 5 \nu$, $\phi_A=0.3$ and $v_A=0.3$.}
\label{variance} 
\end{figure} 
%%%%%%%%%%%%%%%%%%%%%%%%%%%%%%%%%%%%%%%
\subsection{Evolution of coherent states under storage}
\label{sec:storage}
After state preparation, as we will shall see in
Section~\ref{sec:protocol}, Alice needs to
 transfer the ensemble of
dc-SQUIDs to Bob, who then stores it. During the
storage period, the term corresponding to applied
external field is zero and the effective Hamiltonian takes
the form:
\begin{equation}
H = \Omega b^{\dagger}b -\nu(b^{\dagger}b)^2.
\label{eq:hamil_field_off}
\end{equation}
Storing the dc-SQUID under no-dissipation conditions for a
large time $\tau_3$ results in both the phase and the
non-linear term in Eqn.~(\ref{eq:hamil_field_off})
contributing significantly to the evolution of the state.
Let us consider the state at time $t=0$ to be
$|\alpha\rangle=\vert \phi_A+ i v_A\rangle$, 
then the resultant state at a later time $t$ is
\begin{equation}
|\psi(t)\rangle = 
\text{e}^{-i\Omega b^\dagger b t}\text{e}^{i\nu (b^\dagger
b)^2 t}|\alpha\rangle.
\label{eq:resultant_state1}
\end{equation}
In Fock state basis, Eqn.~(\ref{eq:resultant_state1}) becomes
\begin{equation}
|\psi(t)\rangle= e^{-\frac{|\alpha|^2}{2}}
\sum_{n=0}^{\infty} e^{i \nu  n^2 t }\frac{(\alpha e^{-i
\Omega t})^n}{\sqrt{n!}}|n\rangle.
\label{eq:resultant_state2}
\end{equation}
If we engineer the value of $\Omega/\nu$ to be an even
integer, the state~(\ref{eq:resultant_state2}) revives back
to a coherent state $|-\alpha\rangle$ at time $t= \pi/\nu$.
Furthermore, at time $t= 2\pi/\nu$, the state evolves back
to the original state $|\alpha\rangle$. As a general case,
whenever $t = \pi p/\nu q$, where $p<q$ are both mutually
prime, the state~(\ref{eq:resultant_state2}) can be written
as a superposition of coherent states having the same
magnitude $|\alpha|$ but differing in
phase~\cite{Banerji2001,berry1980}:
\begin{eqnarray}
|\psi\left(t= \frac{\pi p}{\nu q}\right)\rangle &=&
\sum_{l=0}^{m-1}c_l^{p,q}|\alpha\text{e}^{-i
\pi\left(\frac{\Omega p}{\nu q}- \frac{2l}{m}\right)}
\rangle,\nonumber \\
{\rm with}&&
c_l^{p,q} = \frac{1}{m}\sum_{r=0}^{m-1}\text{e}^{i \pi
r\left(\frac{pr}{q} - \frac{2l}{m}\right)}
\label{eq:resultant_state3}
\end{eqnarray}
and $m=q$ if at most one of $p$ or $q$ is odd and $m=2q$ if
both are odd.  For the values $p=1$ and $q=2$,
Eqn.~(\ref{eq:resultant_state3}), reduces to  a superposition
of two coherent states as,
\begin{eqnarray}
|\psi\left(t=\frac{\pi}{2\nu} \right)\rangle& =& 
\frac{1}{\sqrt{2}}\left[\text{e}^{i(\pi/4)}|\alpha \text{e}^{-i
\Omega \pi/2 \nu}\rangle\right. \nonumber \\
&&\left.+\text{e}^{-i(\pi/4)}|-\alpha \text{e}^{-i
\Omega \pi/2 \nu}\rangle\right].
\label{eq:superposition_coherent}
\end{eqnarray}
The variance in either quadrature for the 
state~(\ref{eq:resultant_state1}) can be
calculated and is given as
\begin{eqnarray}
\langle(\Delta\Phi)^2\rangle&=&\frac{1}{2}\left[ 
1+2|\alpha|^2+\alpha^2\text{e}^{\left( |\alpha|^2\left(
\gamma^2-1 \right)-2it\xi\right)} 
\right.
\nonumber \\
&&-\beta^2\text{e}^{\left(
|\alpha|^2\left(\gamma^{-2}-1 \right)+2it\xi\right)}
\nonumber \\
&&
\left.
\!\!\!\!\!\!\!\!\!\!\!\!\!\!\!\!\!\!\!\!\!\!\!\!\!\!
+\text{e}^{-2it\zeta}\left(\beta^{*} \text{e}^{\left(
|\alpha|^2\left(\gamma-1\right)\right)}
-\beta\text{e}^{\left(
|\alpha|^2\left(\gamma^{-1}-1\right)+2it\zeta\right)}
\right)^2
\right], 
\label{eq:deltaphi}
\end{eqnarray}
\begin{eqnarray}
\langle(\Delta V)^2\rangle&=&\frac{1}{2}\left[ 
1+ 2 |\alpha|^2-\alpha^2\text{e}^{\left(|\alpha|^2 \left( \gamma^2-1
\right)-2it\xi\right)}
\right.
\nonumber \\
&&
+\beta^2\text{e}^{\left(|\alpha|^2\left(\gamma^{-2}-1
\right)+2it\xi\right)}
\nonumber \\
&&
\!\!\!\!\!\!\!\!\!\!\!\!\!\!\!\!\!\!\!\!\!\!\!\!\!\!
\left.
-\text{e}^{-2it\zeta}\left(
\beta^{*}\text{e}^{\left(|\alpha|^2\left(\gamma-1\right)\right)}+\beta\text{e}^{\left(
|\alpha|^2\left(\gamma^{-1}-1\right)+2it\zeta\right)}
\right)^2
\right],
\label{eq:deltav}
\end{eqnarray}
where $\alpha=(\phi+i v)/\sqrt{2}$, $\xi= \Omega-2\nu$,
$\beta = (v+i \phi)/\sqrt{2}$, $\gamma=\text{e}^{2i\nu t}$
and $\zeta = \Omega-\nu$. A state is said to be squeezed
in quadrature $X\in\lbrace\Phi, V\rbrace$ if its
variance $\langle(\Delta X)^2 \rangle < 1/2$.
Fig.~\ref{variance} shows the plot of variance of both
$\Phi$ and $V$ quadrature with time and their product
showing the appearance and disappearance of squeezing with
time. While the product of variances obey the uncertainty principle
given in Eqn.~(\ref{eq:uncertainty}), the squeezing
appears when one of the variances falls below the coherent
state value.

\begin{figure}
\includegraphics[scale=1]{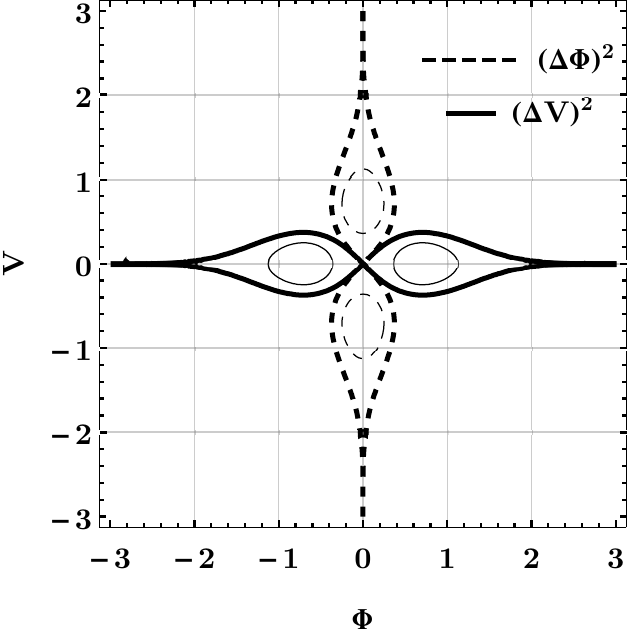}
\caption{Thick-solid and thick-dashed curves represent a
contour of value $0.5$ while thin-solid and thin-dashed
curves represent a contour of value $0.4$. For points inside
contour value $0.5$, state~(\ref{eq:superposition_coherent})
is squeezed in $\Phi$ and $V$ while for points outside this
contour there is no second order squeezing.}
\label{catvariance} 
\end{figure}

As a special case, we analyze squeezing for the
state~(\ref{eq:superposition_coherent}).  For this case
Eqn.~(\ref{eq:deltaphi}) and Eqn.~(\ref{eq:deltav}) yield
\begin{equation}
\langle(\Delta \Phi)^2 \rangle= 
\begin{cases}
\frac{1}{2}+\phi^2-\text{e}^{-2(\phi^2+v^2)}v^2&
\frac{\Omega}{\nu} ~\text{is even}, \\
\frac{1}{2}+v^2-\text{e}^{-2(\phi^2+v^2)}\phi^2            &
\frac{\Omega}{\nu}~ \text{is odd}
\end{cases}
\end{equation}
\begin{equation}
\langle(\Delta V)^2 \rangle= 
\begin{cases}
\frac{1}{2}+v^2-\text{e}^{-2(\phi^2+v^2)}\phi^2&
\frac{\Omega}{\nu} ~\text{is even}, \\
\frac{1}{2}+\phi^2-\text{e}^{-2(\phi^2+v^2)}v^2
& \frac{\Omega}{\nu}~ \text{is odd }
\end{cases}
\end{equation}
Fig.~\ref{catvariance} shows a contour plot for when
$\Omega/\nu$ is even.  The state is found to be squeezed for
only a finite region of $\Phi$ and $V$ inside a contour
value of $0.50$.
%%%%%%%%%%%%%%%%%%%%%%%%%%%%%%
%%%%%%%%%%%%%%%%%%%%%%%%%%%%%%
Therefore, we see that the stored coherent state undergoes
collapse and revival, directly affecting the correlations
between Alice and Bob. In the succeeding subsections, we analyze two
different measurement schemes by Bob and calculate the
average key rate for both the cases.
%%%%%%%%%%%%%%%%%%%%%%%%%%%%%%%%%%%%%%%%%
\section{Protocol}
\label{sec:protocol}
\begin{figure*}
\includegraphics[scale=1]{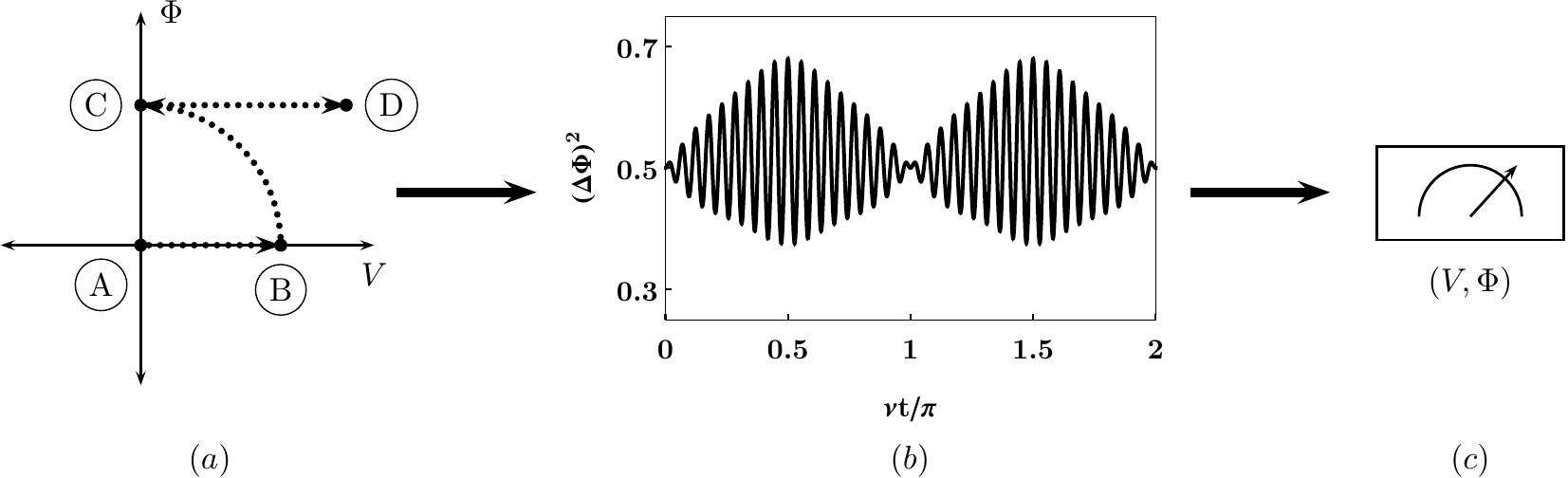} \caption{($a$)
Preparation of the desired coherent state, by displacing the
ground state from A to B followed by a rotation to C and a
subsequent displacement to D. ($b$) Collapse and revival of
the state in the $\Phi$ quadrature when kept in storage for
long durations. We have taken values $\Omega= 20\nu$ for the
purpose of clarity and $\phi_A=0.3$, $v_A=0$. ($c$)
Measurement of the state by Bob by randomly choosing a
quadrature from either $V$ or $\Phi$.} \label{fig:bigfig}
\end{figure*}
In this section we describe our on-demand QKD protocol
based on coherent states prepared on a dc-SQUID.
The key distribution protocol involves following  stages:
\begin{itemize}
\item[S1:]
Alice randomly samples two numbers $\phi_A$ and $v_A$
from two Gaussian distributions with the same variance
$V_AN_0$ and means at $\phi_0$, $v_0$, respectively and
prepares the coherent state $|\phi_A+iv_A\rangle$ on a
dc-SQUID as described in Section~\ref{sec:preparation}.  She
repeats and prepares an ensemble of dc-SQUIDs in coherent
states with $\phi_A$ and $v_A$ chosen randomly from a
Gaussian distribution respectively. 
\item[S2:]
Alice transfers this numbered ensemble of dc-SQUIDS to Bob via a
channel with Gaussian noise. The numbers of SQUID
elements in the ensemble are known to both Alice and Bob.
\item[S3:] 
Bob stores the ensemble after receiving it until a time when
he wants to carry out QKD with Alice. During this storage
period, the states of the members of the SQUIDs ensemble
evolve under the system Hamiltonian and  undergo collapse and
revival as described in Section~\ref{sec:storage}. 
\item[S4:]
At a later time, when the demand for key distribution arises, Bob
performs measurements of one of either voltage $V$
quadrature or flux $\Phi$ quadrature chosen randomly, on
each numbered member of the ensemble. 
\item[S5:]
Afterwards, Bob publicly communicates his choice of
measurement on each of the SQUIDs to Alice. On her side,
Alice only keeps data for each SQUID corresponding to Bob's
measurement quadrature. The correlated data is thus
generated.
\item[S6:] 
The correlated data is transformed into errorless bit
strings by using sliced reconciliation protocols detailed
in~\cite{coh_qkd, Cloning_sec}. Finally they perform privacy
amplification to distill a secure key.
Fig.~\ref{fig:bigfig} illustrates the protocol
schematically. 
\end{itemize}
In the step S4 above Bob can employ two  different
measurements schemes, one in which he does not care about
the precise time for which storage was done and begins to
measure the dc-SQUID quadratures at a time of his choice and
the other in which time stamping is used where when he
begins measurements he uses a specific time for each
dc-SQUID to maximize his correlation with Alice.  For both
the cases we find out the average correlations that can be
shared between Alice and Bob. As we shall see it is possible
to carry out QKD in both the cases while the measurement
scheme with specific time measurements has much higher key
rate.  
%%%%%%%%%%%%%%%%%%%%%%%%%%%%%%%%%%%%%%%%%%%%%%%%%
\subsubsection*{Case 1: Measurement at an arbitrary time}
In this scheme, Bob, after the storage period performs
measurement of either the $\Phi$ or $V$ quadrature without
worrying about the exact time that has elapsed for each
dc-SQUID after its state preparation.  As has been seen in
Section~\ref{sec:storage}, the non-linear evolution of the
state, leads to a periodic variation of correlation between
Alice and Bob as the states undergo periodic collapses and
revivals. Bob's measurement in this case is not sensitive to
this process and gets implement at some random time during
this cycle. Therefore, we assume that the measurement times
are uniformly distributed over the interval $t=0$ and
$t=2\pi/\nu$.  Thus, for this case the relevant correlations
is the time average correlations over this time.
We calculate this correlation and the corresponding
key rate that can be achieved when Bob measures each
dc-SQUID at a random time. To that end, we define a quantity
$C_{AB}(X)$ which quantifies the noise observed by Bob given
the information encoded by Alice when a measurement of $X$
quadrature is performed,
\begin{equation}
\begin{aligned}
C_{AB}(X) &= \langle (X^{B}-X^{A})^2\rangle\\
          &=\langle (X^{B})^2\rangle+(X^{A})^2 - 
             2 X^{A} \langle X^{B}\rangle,
\end{aligned}
\label{eq:corr_avg_bob}
\end{equation}
where the average is taken over the measurement results of
Bob, and $X^A$ and $X^B$ denote the information encoded
by Alice and measurement result of Bob in the $X$ quadrature.

\begin{figure}
\includegraphics[scale=1]{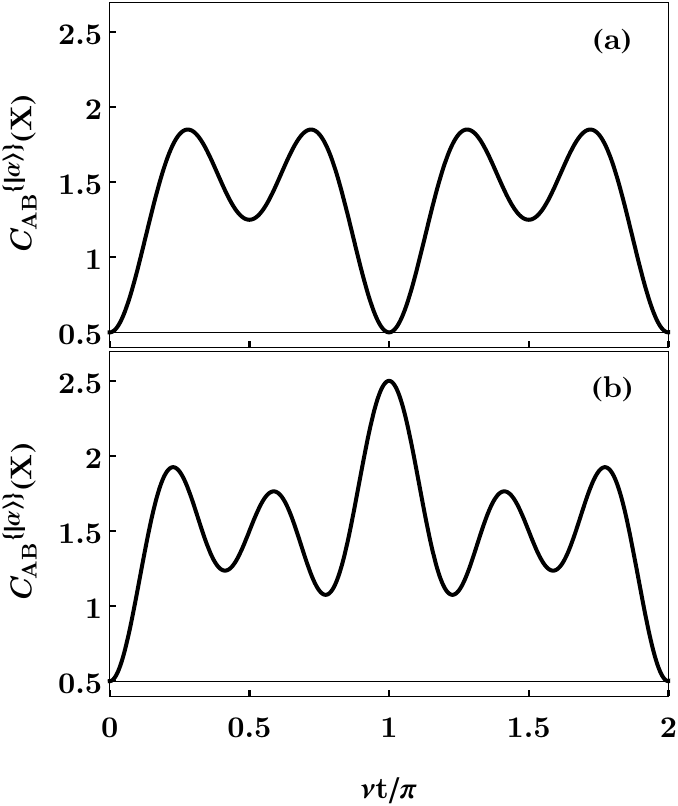}
\caption{($a$) Represents the error in either quadrature for
$\Omega = 5 \nu$. When $\Omega/\nu$ is odd, the
state~(\ref{eq:resultant_state1}) at time $t=0$, $\pi/\nu $
and $ 2\pi/\nu$ is $|\alpha\rangle$ and hence, noise in each
case is equal to $1/2$. ($b$) Represents the error in either
quadrature for $\Omega = 6 \nu$. When $\Omega/\nu$ is even,
the state~(\ref{eq:resultant_state1}) at time $t=0$, and $
2\pi/\nu$ is $|\alpha\rangle$ and noise in both the cases is
equal to $1/2$. However, at time $ \pi/\nu $, the state is
$|-\alpha\rangle$ and hence, the noise is maximum.}
\label{noise} 
\end{figure}

If the state prepared by Alice is $|\alpha\rangle$ upon
which Bob performs a measurement at a random time $t$, the
noise $C_{AB}(X)$ in the two quadratures is found to be
\begin{eqnarray}
C_{AB}(\Phi)
&& = \frac{1}{2}
\left[ 
1+2|\alpha|^2+4\left(\text{Re}\left(\alpha\right)\right)^2
\right.
\nonumber\\
&&\!\!\!\!
-4\text{Re}\left(\alpha\right)\text{e}^{-it\zeta}
\left[\alpha\text{e}^{|\alpha|^2\left(\gamma
-1\right)}+\alpha^*\text{e}^{\left(|\alpha|^2\left(\gamma^{-1}
-1\right)+2it\zeta \right)}\right]
\nonumber \\
&&\!\!\!\!\left. +\alpha^2
\text{e}^{|\alpha|^2\left(\gamma^2-1 \right) -2it\xi}
-\beta^2\text{e}^{\left(
|\alpha|^2\left(\gamma^{-2}-1\right)+2it\xi\right)} 
%\end{aligned}
\right],
\label{eq:cab_phi}
\end{eqnarray}
%%%%%%%%%%%%%%%%%%%%%%%%%%%%%%%%%%%%%%%%
\begin{eqnarray}
C_{AB}(V) &&= \frac{1}{2}\left[ 
%\begin{aligned}
1+2 |\alpha|^2+4\left(\text{Im}\left(\alpha\right)\right)^2
\right.
\nonumber \\
&&\!\!\!\!\!\!\!
-4\text{Im}\left(\alpha\right)\text{e}^{-it\zeta}
\left[\beta\text{e}^{\left(|\alpha|^2\left(\gamma^{-1}-1
\right)+2it\zeta
\right)}+\beta^{*}\text{e}^{|\alpha|^2\left(\gamma
- 1 \right)} \right]
\nonumber \\
&&\!\!\!\!\!\!
\left.
-\alpha^2\text{e}^{\left(|\alpha|^2\left(\gamma^2-1 \right) -2it\xi\right)}
+\beta^2\text{e}^{\left(|\alpha|^2\left(\gamma^{-2}-1\right)+2it\xi\right)}
\right].
\label{eq:cab_v}
\end{eqnarray}
where $\alpha = (\phi_A+i v_A)/\sqrt{2}$, $\xi=
\Omega-2\nu$, $\beta = (v_A+i \phi_A)/\sqrt{2}$,
$\gamma=\text{e}^{2i\nu t}$ and $\zeta = \Omega-\nu$. 
Let us consider that Alice randomly samples $\phi_A$ and
$v_A$ from a Gaussian distribution with variance $V_AN_0=
1/2$ and $\phi_0=v_0=0$. The weighted average noise observed
by Bob in quadrature $X$ over all such states $\{|\alpha
\rangle \}$ randomly chosen by Alice is then given as
\begin{eqnarray}
C_{AB}^{\{|\alpha \rangle \}}(X)&=&
\int_{-\infty}^{\infty}\int_{-\infty}^{\infty}d x_1 d x_2
\frac{1}{\pi}\text{e}^{-(x_1^2 + x_2^2)} C_{AB}(X)
\nonumber \\
&&
\!\!\!\!\!\!\!\!\!\!\!
\!\!\!\!\!\!\!\!\!\!\!
\!\!\!\!\!\!\!\!\!\!\!
=\frac{3}{2}-\frac{9 \cos (t (\nu -\Omega ))-6 \cos (t (\nu
+\Omega ))+\cos (t (3 \nu +\Omega ))}{(5-3 \cos (2 \nu
t))^2}, \nonumber \\
\label{eq:total_noise_analytical}
\end{eqnarray}
where $X$ can be either of the quadratures and the
expression is same for both the quadratures.
Fig.~\ref{noise} shows the plots of $C_{AB}^{\{|\alpha
\rangle \}}(X)$ for odd and even values of $\Omega/\nu$. 

Finally, under the approximation $\Omega\gg\nu$, we average
the noise over the time period $t=0$ to $t=2\pi/\nu$ to get
\begin{equation}
C_{AB}^{\text{T}}(X)=\frac{\nu}{2\pi}
\int_{0}^{\frac{2\pi}{\nu}} dt ~C_{AB}^{\{|\alpha \rangle
\}}(X) = \frac{3}{2}, \label{eq:avg_noise_time}
\end{equation}
and the average correlation between Alice and Bob $I(A:B)$
can be evaluated as
\begin{equation}
I(A:B) = \frac{1}{2}\log_2\left(
1+\frac{V_AN_0}{C_{AB}^{\text{T}}(X)}\right).  
\label{eq:avg_key_time} 
\end{equation}
For $V_AN_0=1/2$ and $C_{AB}^{\text{T}}(X)$ as given by
Eqn.~(\ref{eq:avg_noise_time}) the average correlation comes
out to be $I(A:B) = 0.20$ bits. The protocol thus achieves
its goal of generating data that can be converted
into a key by reconcilliation and privacy amplification.
%%%%%%%%%%%%%%%%%%%%%%%%%%%%%%%%%%%%%%%%
\subsubsection*{Case 2: Measurements at specific times}
%%%%%%%%%%%%%%%%%%%%%%%%%%%
\label{sec:meas_spec_time}
In this scheme, upon the need for QKD, Bob performs the
measurements of $\phi$ or $V$ at specific times chosen to
maximize his correlation with Alice.  These  times
correspond to the times  when the
state~(\ref{eq:resultant_state1}) reverts to a pure coherent
state ($|\alpha\rangle$ or $|-\alpha\rangle$) or an equal
superposition of them as given by
Eqn.~(\ref{eq:superposition_coherent}). The time scale of
oscillations of states are expected to be much smaller
compared to the long storage time. This time keeping has to
be done for each dc-SQUID and therefore we assume that Alice
does precise time stamping for each dc-SQUID. There has to
be synchronization of very precise clocks between Alice and
Bob in order to decide on the measurement times.

\begin{figure}
\includegraphics[scale=1]{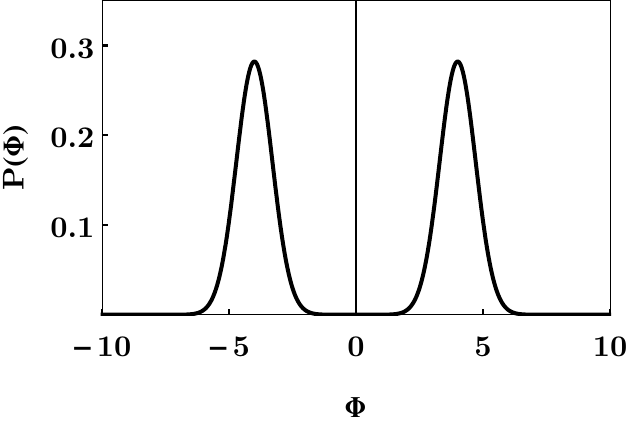}
\caption{The probability distribution of the 
state~(\ref{eq:superposition_coherent}) 
when Alice Gaussian distribution is
centered about $q_0= p_0=4$ and $\Omega=100\nu$.}
\label{fig:catstate} 
\end{figure}

For this scheme, Alice chooses her Gaussian distribution
with variance $V_AN_0=1/2$, but centered around a large
value, e.g. $q_0=p_0=4$. Bob also utilizes a slightly
different scheme for measurements. Apart from measuring at
specified times, he takes the absolute value of the outcomes
he observes.  For both the cases when $\Omega/\nu$ is an odd or an even 
integer and Bob records only the absolute values of his
outcomes, the statistics as observed by him for states at
time $t=0$, $t=\pi/2\nu$, $t=\pi/\nu$, $t=3\pi/2\nu$ and $t=2\pi/\nu$
will have the same mean and variance as Alice's. As is shown
in Fig.~\ref{fig:catstate} he has a bimodal distribution for
the times $t=\pi/2\nu$ and $t=3\pi/2\nu$
which gets folded onto the positive side when we take the
absolute value. For the rest of the times he gets the same statistics as 
Alice's state $|\alpha\rangle$ after taking the absolute value.
 This is the reason that Alice needs to
generate states with large displacement on her side for this
protocol to work.
%%%%%%%%%%%%%%%%%%%%%%%%%%%%%%%%%%%%%%%%%%
%%%%%%%%%%%%%%%%%%%%%%%%%%%%%%%%%%%%%%%%%%%%%%%%%%
The average noise~(\ref{eq:avg_noise_time})
for this measurement scheme by Bob at these specific time
intervals is only the vacuum noise $N_0=1/2$. 
The average correlations can then be computed as
\begin{equation}
I(A:B) = \frac{1}{2}\log_2\left(1+\frac{V_AN_0}{N_0}\right)
= 0.50~\text{bits}.
\label{eq:avg_key_spec_time}
\end{equation}
As has been emphasized, in order to achieve the
aforementioned correlations, it is essential to perform
measurements at the precisely specified times. Since Bob
will be storing these SQUIDs for a long time, it is also
required to keep an accurate track of time for each and
every SQUID.  From an experimental point of view we can
assume that the collapse and revival time period lies in the
range of a few hundred microseconds $(10^{-4}$s)
corresponding to $\Omega\approx 10^{4}$Hz and $\Omega/\nu =
100$.  Commercially available atomic clocks can keep track
of time upto an error of $0.1\mu$s per day amounting to
$1000$ days until the error becomes significant.
Furthermore, with the current computer processors in the
range of GHz, it is possible to perform precise measurements
with a resolution of nanoseconds. Therefore the
correlations~(\ref{eq:avg_key_spec_time}) are achievable
with the current technology for storage times of
approximately $30$ months.
%%%%%%%%%%%%%%%%%%%%%%%%%%%%%%%%%%%%%%%%%%%%%%
\section{Security}
\label{sec:security}
We analyze security of the protocol for both the cases (with
and without time stamping) under an assumption of individual
attacks by an eavesdropper, Eve, where she ends up
introducing Gaussian noise in the dc-SQUIDs.

The security proof of the protocol is quite similar to the
one for continuous variable QKD based on coherent
states~\cite{coh_qkd}. We assume that Eve has gained access
to the ensemble of SQUIDs prepared by Alice while being
transported through a channel with Gaussian noise.  According
to a result in~\cite{qkd_sec_nocloning} based on the
no-cloning principle and incompatibility of measurements, 
if the noise added on Bob's side is
$\chi N_0$, then the minimum added noise on Eve's side is
$\chi^{-1}N_0$. In the presence of this noise, for the
scheme where Bob is performing measurements at arbitrary
times, the secure key rate can be found out as
\begin{equation}
\begin{aligned}
\Delta I &= I(A:B)-I(A:E)\\
 &=
\frac{1}{2}\log_2\left(\frac{\left(V_A+\chi\right)N_0+C_{AB}^T(X_i)}{\chi
N_0+C_{AB}^T(X)}\right)\\
&\phantom{=}-\frac{1}{2}\log_2\left(\frac{\left(1+V_A\chi\right)N_0+
C_{AE}^T(X)\chi}{C_{AE}^T(X)\chi+N_0}\right),
\end{aligned}
\label{eq:sec_key_arb}
\end{equation}
where $C_{AE}^T(X)$ is the average
noise in the correlations of Alice and Eve. The strategy
employed by Eve should be such that her average noise is
minimized. One of the best strategies that accomplishes this
is the one employed by Bob when measuring at specified
times. The collapse and revival phenomenon also ensures that
the minimum noise in Eve's data cannot be less than the
vacuum noise $N_0$. In order to achieve a positive secure
key rate from Eqn.~(\ref{eq:sec_key_arb}), it is required
that $\chi<1$. For the case when Bob is also measuring at
specified times, the condition for secure key rate can be
found by putting $C_{AB}^T(X) = N_0 = 1/2$ in
Eqn.~(\ref{eq:sec_key_arb}). For this case we again arrive
at the condition $\chi<1$.
Therefore in both the cases, observation of external noise 
$\chi\geq1$ implies that either Eve is present or there has
been too much noise and the key cannot be distilled. 
%%%%%%%%%%%%%%%%%%%%%%%%%%%%%%%%%%%%%%%%%%%%%%%%%%%%%
\section{Conclusion}
\label{sec:conc}
In this paper we explored the possibility of carrying out
QKD with a non-photonic quantum system.  The protocol
involves preparation of coherent states on a dc-SQUID by
Alice, their  transportation to Bob's location, subsequent
storage under no dissipation conditions~\cite{dissipationless1,dissipationless2} in Bob's lab, and
finally quadrature measurements and classical information
processing between Alice and Bob.  We exploited the
longevity of superconducting coherent states to perform
on-demand secure quantum key distribution where Bob stores
the dc-SQUIDS prepared in coherent states and carries out
measurements only when the key is required.  Given the
Hamiltonian of the dc-SQUID, during the storage period the
correlations between Alice and Bob undergo collapses and
revivals. This motivated us to design two variants of the
protocol, one without time stamping and other with time
stamping. While the protocol with time stamping gives a
higher key rate of $0.5$ bits, it is more difficult to carry
out.  What is noteworthy is that even the simpler protocol,
where Alice and Bob do not perform any time stamping, has a
reasonable key rate of $0.2$ bits.  Our  protocol enjoys
several advantages over standard photon based QKD protocols,
the foremost being the storage possibility wherein Alice and
Bob can introduce a delay of years between the exchange of
states and key generation which can be done on demand.
Secondly, the on-demand QKD protocol acts as a bridge
between SQUIDs, which have found extensive application in
quantum information processing in recent times and quantum
communication~\cite{superconducting_bits,squid_algo,squid_comp}.
This opens up several new and interesting avenues of
research in application of other condensed matter concepts
to quantum communication.  In particular it would be
interesting to see how entangled states of SQUIDs can be
prepared for application in entanglement assisted QKD.  We
hope that this work will generate interest in inventing more
such protocols and also in experimentally building such a
QKD system.  
%%%%%%%%%%%%%%%%%%%%%%%%%%%%%%%%%%%%%%%%%%%%%%%%%%%%%%%%
\begin{acknowledgements}
J.S. acknowledges UGC India for financial support.
A and C. K. acknowledge funding from DST India
under Grant No. EMR/2014/000297.
\end{acknowledgements}
%merlin.mbs apsrev4-1.bst 2010-07-25 4.21a (PWD, AO, DPC) hacked
%Control: key (0)
%Control: author (8) initials jnrlst
%Control: editor formatted (1) identically to author
%Control: production of article title (-1) disabled
%Control: page (0) single
%Control: year (1) truncated
%Control: production of eprint (0) enabled
%

%%\bibliography{on_demand_qkd}

\begin{thebibliography}{35}%
\makeatletter
\providecommand \@ifxundefined [1]{%
 \@ifx{#1\undefined}
}%
\providecommand \@ifnum [1]{%
 \ifnum #1\expandafter \@firstoftwo
 \else \expandafter \@secondoftwo
 \fi
}%
\providecommand \@ifx [1]{%
 \ifx #1\expandafter \@firstoftwo
 \else \expandafter \@secondoftwo
 \fi
}%
\providecommand \natexlab [1]{#1}%
\providecommand \enquote  [1]{``#1''}%
\providecommand \bibnamefont  [1]{#1}%
\providecommand \bibfnamefont [1]{#1}%
\providecommand \citenamefont [1]{#1}%
\providecommand \href@noop [0]{\@secondoftwo}%
\providecommand \href [0]{\begingroup \@sanitize@url \@href}%
\providecommand \@href[1]{\@@startlink{#1}\@@href}%
\providecommand \@@href[1]{\endgroup#1\@@endlink}%
\providecommand \@sanitize@url [0]{\catcode `\\12\catcode `\$12\catcode
  `\&12\catcode `\#12\catcode `\^12\catcode `\_12\catcode `\%12\relax}%
\providecommand \@@startlink[1]{}%
\providecommand \@@endlink[0]{}%
\providecommand \url  [0]{\begingroup\@sanitize@url \@url }%
\providecommand \@url [1]{\endgroup\@href {#1}{\urlprefix }}%
\providecommand \urlprefix  [0]{URL }%
\providecommand \Eprint [0]{\href }%
\providecommand \doibase [0]{http://dx.doi.org/}%
\providecommand \selectlanguage [0]{\@gobble}%
\providecommand \bibinfo  [0]{\@secondoftwo}%
\providecommand \bibfield  [0]{\@secondoftwo}%
\providecommand \translation [1]{[#1]}%
\providecommand \BibitemOpen [0]{}%
\providecommand \bibitemStop [0]{}%
\providecommand \bibitemNoStop [0]{.\EOS\space}%
\providecommand \EOS [0]{\spacefactor3000\relax}%
\providecommand \BibitemShut  [1]{\csname bibitem#1\endcsname}%
\let\auto@bib@innerbib\@empty
%</preamble>
\bibitem [{\citenamefont {Bennett}\ and\ \citenamefont
  {Brassard}(1984)}]{BB84}%
  \BibitemOpen
  \bibfield  {author} {\bibinfo {author} {\bibfnamefont {C.~H.}\ \bibnamefont
  {Bennett}}\ and\ \bibinfo {author} {\bibfnamefont {G.}~\bibnamefont
  {Brassard}},\ }in\ \href@noop {} {{\bibinfo {booktitle}
{\it Proceedings of
  the IEEE International Conference on Computers, Systems and Signal
  Processing}}}\ (\bibinfo  {publisher} {IEEE Press},\ \bibinfo {address} {New
  York},\ \bibinfo {year} {1984})\ pp.\ \bibinfo {pages} {175--179}\BibitemShut
  {NoStop}%
\bibitem [{\citenamefont {Bennett}(1992)}]{B92}%
  \BibitemOpen
  \bibfield  {author} {\bibinfo {author} {\bibfnamefont {C.~H.}\ \bibnamefont
  {Bennett}},\ }\href {\doibase 10.1103/PhysRevLett.68.3121} {\bibfield
  {journal} {\bibinfo  {journal} {Phys. Rev. Lett.}\ }\textbf {\bibinfo
  {volume} {68}},\ \bibinfo {pages} {3121} (\bibinfo {year}
  {1992})}\BibitemShut {NoStop}%
\bibitem [{\citenamefont {Gisin}\ \emph {et~al.}(2002)\citenamefont {Gisin},
  \citenamefont {Ribordy}, \citenamefont {Tittel},\ and\ \citenamefont
  {Zbinden}}]{Crypt_review}%
  \BibitemOpen
  \bibfield  {author} {\bibinfo {author} {\bibfnamefont {N.}~\bibnamefont
  {Gisin}}, \bibinfo {author} {\bibfnamefont {G.}~\bibnamefont {Ribordy}},
  \bibinfo {author} {\bibfnamefont {W.}~\bibnamefont {Tittel}}, \ and\ \bibinfo
  {author} {\bibfnamefont {H.}~\bibnamefont {Zbinden}},\ }\href {\doibase
  10.1103/RevModPhys.74.145} {\bibfield  {journal} {\bibinfo  {journal} {Rev.
  Mod. Phys.}\ }\textbf {\bibinfo {volume} {74}},\ \bibinfo {pages} {145}
  (\bibinfo {year} {2002})}\BibitemShut {NoStop}%
\bibitem [{\citenamefont {Shor}\ and\ \citenamefont
  {Preskill}(2000)}]{Shor_Pres_2000}%
  \BibitemOpen
  \bibfield  {author} {\bibinfo {author} {\bibfnamefont {P.~W.}\ \bibnamefont
  {Shor}}\ and\ \bibinfo {author} {\bibfnamefont {J.}~\bibnamefont
  {Preskill}},\ }\href {\doibase 10.1103/PhysRevLett.85.441} {\bibfield
  {journal} {\bibinfo  {journal} {Phys. Rev. Lett.}\ }\textbf {\bibinfo
  {volume} {85}},\ \bibinfo {pages} {441} (\bibinfo {year} {2000})}\BibitemShut
  {NoStop}%
\bibitem [{\citenamefont {Branciard}\ \emph {et~al.}(2005)\citenamefont
  {Branciard}, \citenamefont {Gisin}, \citenamefont {Kraus},\ and\
  \citenamefont {Scarani}}]{Sec_BB84}%
  \BibitemOpen
  \bibfield  {author} {\bibinfo {author} {\bibfnamefont {C.}~\bibnamefont
  {Branciard}}, \bibinfo {author} {\bibfnamefont {N.}~\bibnamefont {Gisin}},
  \bibinfo {author} {\bibfnamefont {B.}~\bibnamefont {Kraus}}, \ and\ \bibinfo
  {author} {\bibfnamefont {V.}~\bibnamefont {Scarani}},\ }\href {\doibase
  10.1103/PhysRevA.72.032301} {\bibfield  {journal} {\bibinfo  {journal} {Phys.
  Rev. A}\ }\textbf {\bibinfo {volume} {72}},\ \bibinfo {pages} {032301}
  (\bibinfo {year} {2005})}\BibitemShut {NoStop}%
\bibitem [{\citenamefont {Ekert}(1991)}]{E91}%
  \BibitemOpen
  \bibfield  {author} {\bibinfo {author} {\bibfnamefont {A.~K.}\ \bibnamefont
  {Ekert}},\ }\href {\doibase 10.1103/PhysRevLett.67.661} {\bibfield  {journal}
  {\bibinfo  {journal} {Phys. Rev. Lett.}\ }\textbf {\bibinfo {volume} {67}},\
  \bibinfo {pages} {661} (\bibinfo {year} {1991})}\BibitemShut {NoStop}%
\bibitem [{\citenamefont {Ac\'{\i}n}\ \emph {et~al.}(2006)\citenamefont
  {Ac\'{\i}n}, \citenamefont {Gisin},\ and\ \citenamefont
  {Masanes}}]{Acin_sec}%
  \BibitemOpen
  \bibfield  {author} {\bibinfo {author} {\bibfnamefont {A.}~\bibnamefont
  {Ac\'{\i}n}}, \bibinfo {author} {\bibfnamefont {N.}~\bibnamefont {Gisin}}, \
  and\ \bibinfo {author} {\bibfnamefont {L.}~\bibnamefont {Masanes}},\ }\href
  {\doibase 10.1103/PhysRevLett.97.120405} {\bibfield  {journal} {\bibinfo
  {journal} {Phys. Rev. Lett.}\ }\textbf {\bibinfo {volume} {97}},\ \bibinfo
  {pages} {120405} (\bibinfo {year} {2006})}\BibitemShut {NoStop}%
\bibitem [{\citenamefont {Cerf}\ \emph {et~al.}(2002)\citenamefont {Cerf},
  \citenamefont {Iblisdir},\ and\ \citenamefont {Van~Assche}}]{Cloning_sec}%
  \BibitemOpen
  \bibfield  {author} {\bibinfo {author} {\bibfnamefont {N.}~\bibnamefont
  {Cerf}}, \bibinfo {author} {\bibfnamefont {S.}~\bibnamefont {Iblisdir}}, \
  and\ \bibinfo {author} {\bibfnamefont {G.}~\bibnamefont {Van~Assche}},\
  }\href {\doibase 10.1140/epjd/e20020025} {\bibfield  {journal} {\bibinfo
  {journal} {The European Physical Journal D - Atomic, Molecular, Optical and
  Plasma Physics}\ }\textbf {\bibinfo {volume} {18}},\ \bibinfo {pages} {211}
  (\bibinfo {year} {2002})}\BibitemShut {NoStop}%
\bibitem [{\citenamefont {Ralph}(1999)}]{cv_qkd_ralph}%
  \BibitemOpen
  \bibfield  {author} {\bibinfo {author} {\bibfnamefont {T.~C.}\ \bibnamefont
  {Ralph}},\ }\href {\doibase 10.1103/PhysRevA.61.010303} {\bibfield  {journal}
  {\bibinfo  {journal} {Phys. Rev. A}\ }\textbf {\bibinfo {volume} {61}},\
  \bibinfo {pages} {010303} (\bibinfo {year} {1999})}\BibitemShut {NoStop}%
\bibitem [{\citenamefont {Paw\l{}owski}(2010)}]{Sec_monogamy}%
  \BibitemOpen
  \bibfield  {author} {\bibinfo {author} {\bibfnamefont {M.}~\bibnamefont
  {Paw\l{}owski}},\ }\href {\doibase 10.1103/PhysRevA.82.032313} {\bibfield
  {journal} {\bibinfo  {journal} {Phys. Rev. A}\ }\textbf {\bibinfo {volume}
  {82}},\ \bibinfo {pages} {032313} (\bibinfo {year} {2010})}\BibitemShut
  {NoStop}%
\bibitem [{\citenamefont {Singh}\ \emph {et~al.}(2017)\citenamefont {Singh},
  \citenamefont {Bharti},\ and\ \citenamefont {Arvind}}]{jask_mono}%
  \BibitemOpen
  \bibfield  {author} {\bibinfo {author} {\bibfnamefont {J.}~\bibnamefont
  {Singh}}, \bibinfo {author} {\bibfnamefont {K.}~\bibnamefont {Bharti}}, \
  and\ \bibinfo {author} {\bibnamefont {Arvind}},\ }\href {\doibase
  10.1103/PhysRevA.95.062333} {\bibfield  {journal} {\bibinfo  {journal} {Phys.
  Rev. A}\ }\textbf {\bibinfo {volume} {95}},\ \bibinfo {pages} {062333}
  (\bibinfo {year} {2017})}\BibitemShut {NoStop}%
\bibitem [{\citenamefont {Scarani}\ \emph {et~al.}(2009)\citenamefont
  {Scarani}, \citenamefont {Bechmann-Pasquinucci}, \citenamefont {Cerf},
  \citenamefont {Du\ifmmode~\check{s}\else \v{s}\fi{}ek}, \citenamefont
  {L\"utkenhaus},\ and\ \citenamefont {Peev}}]{Sec_qkd}%
  \BibitemOpen
  \bibfield  {author} {\bibinfo {author} {\bibfnamefont {V.}~\bibnamefont
  {Scarani}}, \bibinfo {author} {\bibfnamefont {H.}~\bibnamefont
  {Bechmann-Pasquinucci}}, \bibinfo {author} {\bibfnamefont {N.~J.}\
  \bibnamefont {Cerf}}, \bibinfo {author} {\bibfnamefont {M.}~\bibnamefont
  {Du\ifmmode~\check{s}\else \v{s}\fi{}ek}}, \bibinfo {author} {\bibfnamefont
  {N.}~\bibnamefont {L\"utkenhaus}}, \ and\ \bibinfo {author} {\bibfnamefont
  {M.}~\bibnamefont {Peev}},\ }\href {\doibase 10.1103/RevModPhys.81.1301}
  {\bibfield  {journal} {\bibinfo  {journal} {Rev. Mod. Phys.}\ }\textbf
  {\bibinfo {volume} {81}},\ \bibinfo {pages} {1301} (\bibinfo {year}
  {2009})}\BibitemShut {NoStop}%
\bibitem [{\citenamefont {Ralph}(2000)}]{sec_cv_ralph}%
  \BibitemOpen
  \bibfield  {author} {\bibinfo {author} {\bibfnamefont {T.~C.}\ \bibnamefont
  {Ralph}},\ }\href {\doibase 10.1103/PhysRevA.62.062306} {\bibfield  {journal}
  {\bibinfo  {journal} {Phys. Rev. A}\ }\textbf {\bibinfo {volume} {62}},\
  \bibinfo {pages} {062306} (\bibinfo {year} {2000})}\BibitemShut {NoStop}%
\bibitem [{\citenamefont {Grosshans}\ and\ \citenamefont
  {Grangier}(2002)}]{coh_qkd}%
  \BibitemOpen
  \bibfield  {author} {\bibinfo {author} {\bibfnamefont {F.}~\bibnamefont
  {Grosshans}}\ and\ \bibinfo {author} {\bibfnamefont {P.}~\bibnamefont
  {Grangier}},\ }\href {\doibase 10.1103/PhysRevLett.88.057902} {\bibfield
  {journal} {\bibinfo  {journal} {Phys. Rev. Lett.}\ }\textbf {\bibinfo
  {volume} {88}},\ \bibinfo {pages} {057902} (\bibinfo {year}
  {2002})}\BibitemShut {NoStop}%
\bibitem [{\citenamefont {Grosshans}\ \emph {et~al.}(2003)\citenamefont
  {Grosshans}, \citenamefont {Van~Assche}, \citenamefont {Wenger},
  \citenamefont {Brouri}, \citenamefont {Cerf},\ and\ \citenamefont
  {Grangier}}]{expt_cv_qkd}%
  \BibitemOpen
  \bibfield  {author} {\bibinfo {author} {\bibfnamefont {F.}~\bibnamefont
  {Grosshans}}, \bibinfo {author} {\bibfnamefont {G.}~\bibnamefont
  {Van~Assche}}, \bibinfo {author} {\bibfnamefont {J.}~\bibnamefont {Wenger}},
  \bibinfo {author} {\bibfnamefont {R.}~\bibnamefont {Brouri}}, \bibinfo
  {author} {\bibfnamefont {N.~J.}\ \bibnamefont {Cerf}}, \ and\ \bibinfo
  {author} {\bibfnamefont {P.}~\bibnamefont {Grangier}},\ }\href
  {http://dx.doi.org/10.1038/nature01289} {\bibfield  {journal} {\bibinfo
  {journal} {Nature}\ }\textbf {\bibinfo {volume} {421}},\ \bibinfo {pages}
  {238 EP } (\bibinfo {year} {2003})}\BibitemShut {NoStop}%
\bibitem [{\citenamefont {Takesue}\ \emph {et~al.}(2007)\citenamefont
  {Takesue}, \citenamefont {Nam}, \citenamefont {Zhang}, \citenamefont
  {Hadfield}, \citenamefont {Honjo}, \citenamefont {Tamaki},\ and\
  \citenamefont {Yamamoto}}]{expt_qkd1}%
  \BibitemOpen
  \bibfield  {author} {\bibinfo {author} {\bibfnamefont {H.}~\bibnamefont
  {Takesue}}, \bibinfo {author} {\bibfnamefont {S.~W.}\ \bibnamefont {Nam}},
  \bibinfo {author} {\bibfnamefont {Q.}~\bibnamefont {Zhang}}, \bibinfo
  {author} {\bibfnamefont {R.~H.}\ \bibnamefont {Hadfield}}, \bibinfo {author}
  {\bibfnamefont {T.}~\bibnamefont {Honjo}}, \bibinfo {author} {\bibfnamefont
  {K.}~\bibnamefont {Tamaki}}, \ and\ \bibinfo {author} {\bibfnamefont
  {Y.}~\bibnamefont {Yamamoto}},\ }\href
  {http://dx.doi.org/10.1038/nphoton.2007.75} {\bibfield  {journal} {\bibinfo
  {journal} {Nature Photonics}\ }\textbf {\bibinfo {volume} {1}},\ \bibinfo
  {pages} {343 EP } (\bibinfo {year} {2007})}\BibitemShut {NoStop}%
\bibitem [{\citenamefont {Fedrizzi}\ \emph {et~al.}(2009)\citenamefont
  {Fedrizzi}, \citenamefont {Ursin}, \citenamefont {Herbst}, \citenamefont
  {Nespoli}, \citenamefont {Prevedel}, \citenamefont {Scheidl}, \citenamefont
  {Tiefenbacher}, \citenamefont {Jennewein},\ and\ \citenamefont
  {Zeilinger}}]{expt_qkd2}%
  \BibitemOpen
  \bibfield  {author} {\bibinfo {author} {\bibfnamefont {A.}~\bibnamefont
  {Fedrizzi}}, \bibinfo {author} {\bibfnamefont {R.}~\bibnamefont {Ursin}},
  \bibinfo {author} {\bibfnamefont {T.}~\bibnamefont {Herbst}}, \bibinfo
  {author} {\bibfnamefont {M.}~\bibnamefont {Nespoli}}, \bibinfo {author}
  {\bibfnamefont {R.}~\bibnamefont {Prevedel}}, \bibinfo {author}
  {\bibfnamefont {T.}~\bibnamefont {Scheidl}}, \bibinfo {author} {\bibfnamefont
  {F.}~\bibnamefont {Tiefenbacher}}, \bibinfo {author} {\bibfnamefont
  {T.}~\bibnamefont {Jennewein}}, \ and\ \bibinfo {author} {\bibfnamefont
  {A.}~\bibnamefont {Zeilinger}},\ }\href {http://dx.doi.org/10.1038/nphys1255}
  {\bibfield  {journal} {\bibinfo  {journal} {Nature Physics}\ }\textbf
  {\bibinfo {volume} {5}},\ \bibinfo {pages} {389 EP } (\bibinfo {year}
  {2009})}\BibitemShut {NoStop}%
\bibitem [{\citenamefont {Jouguet}\ \emph {et~al.}(2013)\citenamefont
  {Jouguet}, \citenamefont {Kunz-Jacques}, \citenamefont {Leverrier},
  \citenamefont {Grangier},\ and\ \citenamefont {Diamanti}}]{expt_qkd3}%
  \BibitemOpen
  \bibfield  {author} {\bibinfo {author} {\bibfnamefont {P.}~\bibnamefont
  {Jouguet}}, \bibinfo {author} {\bibfnamefont {S.}~\bibnamefont
  {Kunz-Jacques}}, \bibinfo {author} {\bibfnamefont {A.}~\bibnamefont
  {Leverrier}}, \bibinfo {author} {\bibfnamefont {P.}~\bibnamefont {Grangier}},
  \ and\ \bibinfo {author} {\bibfnamefont {E.}~\bibnamefont {Diamanti}},\
  }\href {http://dx.doi.org/10.1038/nphoton.2013.63} {\bibfield  {journal}
  {\bibinfo  {journal} {Nature Photonics}\ }\textbf {\bibinfo {volume} {7}},\
  \bibinfo {pages} {378 EP } (\bibinfo {year} {2013})}\BibitemShut {NoStop}%
\bibitem [{\citenamefont {Takenaka}\ \emph {et~al.}(2017)\citenamefont
  {Takenaka}, \citenamefont {Carrasco-Casado}, \citenamefont {Fujiwara},
  \citenamefont {Kitamura}, \citenamefont {Sasaki},\ and\ \citenamefont
  {Toyoshima}}]{expt_qkd4}%
  \BibitemOpen
  \bibfield  {author} {\bibinfo {author} {\bibfnamefont {H.}~\bibnamefont
  {Takenaka}}, \bibinfo {author} {\bibfnamefont {A.}~\bibnamefont
  {Carrasco-Casado}}, \bibinfo {author} {\bibfnamefont {M.}~\bibnamefont
  {Fujiwara}}, \bibinfo {author} {\bibfnamefont {M.}~\bibnamefont {Kitamura}},
  \bibinfo {author} {\bibfnamefont {M.}~\bibnamefont {Sasaki}}, \ and\ \bibinfo
  {author} {\bibfnamefont {M.}~\bibnamefont {Toyoshima}},\ }\href
  {http://dx.doi.org/10.1038/nphoton.2017.107} {\bibfield  {journal} {\bibinfo
  {journal} {Nature Photonics}\ }\textbf {\bibinfo {volume} {11}},\ \bibinfo
  {pages} {502 EP } (\bibinfo {year} {2017})}\BibitemShut {NoStop}%
\bibitem [{\citenamefont {Joshi}(2000)}]{junction}%
  \BibitemOpen
  \bibfield  {author} {\bibinfo {author} {\bibfnamefont {A.}~\bibnamefont
  {Joshi}},\ }\href {\doibase https://doi.org/10.1016/S0375-9601(00)00313-3}
  {\bibfield  {journal} {\bibinfo  {journal} {Physics Letters A}\ }\textbf
  {\bibinfo {volume} {270}},\ \bibinfo {pages} {249 } (\bibinfo {year}
  {2000})}\BibitemShut {NoStop}%
\bibitem [{\citenamefont {Everitt}\ \emph {et~al.}(2001)\citenamefont
  {Everitt}, \citenamefont {Stiffell}, \citenamefont {Clark}, \citenamefont
  {Vourdas}, \citenamefont {Ralph}, \citenamefont {Prance},\ and\ \citenamefont
  {Prance}}]{squid_ring}%
  \BibitemOpen
  \bibfield  {author} {\bibinfo {author} {\bibfnamefont {M.~J.}\ \bibnamefont
  {Everitt}}, \bibinfo {author} {\bibfnamefont {P.}~\bibnamefont {Stiffell}},
  \bibinfo {author} {\bibfnamefont {T.~D.}\ \bibnamefont {Clark}}, \bibinfo
  {author} {\bibfnamefont {A.}~\bibnamefont {Vourdas}}, \bibinfo {author}
  {\bibfnamefont {J.~F.}\ \bibnamefont {Ralph}}, \bibinfo {author}
  {\bibfnamefont {H.}~\bibnamefont {Prance}}, \ and\ \bibinfo {author}
  {\bibfnamefont {R.~J.}\ \bibnamefont {Prance}},\ }\href {\doibase
  10.1103/PhysRevB.63.144530} {\bibfield  {journal} {\bibinfo  {journal} {Phys.
  Rev. B}\ }\textbf {\bibinfo {volume} {63}},\ \bibinfo {pages} {144530}
  (\bibinfo {year} {2001})}\BibitemShut {NoStop}%
\bibitem [{\citenamefont {Fagaly}(2006)}]{squid}%
  \BibitemOpen
  \bibfield  {author} {\bibinfo {author} {\bibfnamefont {R.~L.}\ \bibnamefont
  {Fagaly}},\ }\href {\doibase 10.1063/1.2354545} {\bibfield  {journal}
  {\bibinfo  {journal} {Review of Scientific Instruments}\ }\textbf {\bibinfo
  {volume} {77}},\ \bibinfo {pages} {101101} (\bibinfo {year}
  {2006})}\BibitemShut {NoStop}%
\bibitem [{\citenamefont {Zou}\ and\ \citenamefont {Shao}(1999)}]{squid_hamil}%
  \BibitemOpen
  \bibfield  {author} {\bibinfo {author} {\bibfnamefont {J.}~\bibnamefont
  {Zou}}\ and\ \bibinfo {author} {\bibfnamefont {B.}~\bibnamefont {Shao}},\
  }\href {\doibase 10.1142/S021797929900076X} {\bibfield  {journal} {\bibinfo
  {journal} {International Journal of Modern Physics B}\ }\textbf {\bibinfo
  {volume} {13}},\ \bibinfo {pages} {917} (\bibinfo {year} {1999})}\BibitemShut
  {NoStop}%
\bibitem [{\citenamefont {Zou}\ and\ \citenamefont
  {Shao}(2001)}]{squid_hamil_diss}%
  \BibitemOpen
  \bibfield  {author} {\bibinfo {author} {\bibfnamefont {J.}~\bibnamefont
  {Zou}}\ and\ \bibinfo {author} {\bibfnamefont {B.}~\bibnamefont {Shao}},\
  }\href {\doibase 10.1103/PhysRevB.64.024511} {\bibfield  {journal} {\bibinfo
  {journal} {Phys. Rev. B}\ }\textbf {\bibinfo {volume} {64}},\ \bibinfo
  {pages} {024511} (\bibinfo {year} {2001})}\BibitemShut {NoStop}%
\bibitem [{\citenamefont {Friedman}\ \emph {et~al.}(2000)\citenamefont
  {Friedman}, \citenamefont {Patel}, \citenamefont {Chen}, \citenamefont
  {Tolpygo},\ and\ \citenamefont {Lukens}}]{Friedman2000}%
  \BibitemOpen
  \bibfield  {author} {\bibinfo {author} {\bibfnamefont {J.~R.}\ \bibnamefont
  {Friedman}}, \bibinfo {author} {\bibfnamefont {V.}~\bibnamefont {Patel}},
  \bibinfo {author} {\bibfnamefont {W.}~\bibnamefont {Chen}}, \bibinfo {author}
  {\bibfnamefont {S.~K.}\ \bibnamefont {Tolpygo}}, \ and\ \bibinfo {author}
  {\bibfnamefont {J.~E.}\ \bibnamefont {Lukens}},\ }\href
  {http://dx.doi.org/10.1038/35017505} {\bibfield  {journal} {\bibinfo
  {journal} {Nature}\ }\textbf {\bibinfo {volume} {406}},\ \bibinfo {pages} {43
  EP } (\bibinfo {year} {2000})}\BibitemShut {NoStop}%
\bibitem [{\citenamefont {Mailly}\ \emph {et~al.}(1993)\citenamefont {Mailly},
  \citenamefont {Chapelier},\ and\ \citenamefont {Benoit}}]{dissipationless1}%
  \BibitemOpen
  \bibfield  {author} {\bibinfo {author} {\bibfnamefont {D.}~\bibnamefont
  {Mailly}}, \bibinfo {author} {\bibfnamefont {C.}~\bibnamefont {Chapelier}}, \
  and\ \bibinfo {author} {\bibfnamefont {A.}~\bibnamefont {Benoit}},\ }\href
  {\doibase 10.1103/PhysRevLett.70.2020} {\bibfield  {journal} {\bibinfo
  {journal} {Phys. Rev. Lett.}\ }\textbf {\bibinfo {volume} {70}},\ \bibinfo
  {pages} {2020} (\bibinfo {year} {1993})}\BibitemShut {NoStop}%
\bibitem [{\citenamefont {L\'evy}\ \emph {et~al.}(1990)\citenamefont {L\'evy},
  \citenamefont {Dolan}, \citenamefont {Dunsmuir},\ and\ \citenamefont
  {Bouchiat}}]{dissipationless2}%
  \BibitemOpen
  \bibfield  {author} {\bibinfo {author} {\bibfnamefont {L.~P.}\ \bibnamefont
  {L\'evy}}, \bibinfo {author} {\bibfnamefont {G.}~\bibnamefont {Dolan}},
  \bibinfo {author} {\bibfnamefont {J.}~\bibnamefont {Dunsmuir}}, \ and\
  \bibinfo {author} {\bibfnamefont {H.}~\bibnamefont {Bouchiat}},\ }\href
  {\doibase 10.1103/PhysRevLett.64.2074} {\bibfield  {journal} {\bibinfo
  {journal} {Phys. Rev. Lett.}\ }\textbf {\bibinfo {volume} {64}},\ \bibinfo
  {pages} {2074} (\bibinfo {year} {1990})}\BibitemShut {NoStop}%
\bibitem [{\citenamefont {Shannon}(1948)}]{shannon}%
  \BibitemOpen
  \bibfield  {author} {\bibinfo {author} {\bibfnamefont {C.~E.}\ \bibnamefont
  {Shannon}},\ }\href {\doibase 10.1002/j.1538-7305.1948.tb01338.x} {\bibfield
  {journal} {\bibinfo  {journal} {Bell System Technical Journal}\ }\textbf
  {\bibinfo {volume} {27}},\ \bibinfo {pages} {379} (\bibinfo {year}
  {1948})}\BibitemShut {NoStop}%
\bibitem [{\citenamefont {Grosshans}\ and\ \citenamefont
  {Grangier}(2001)}]{qkd_sec_nocloning}%
  \BibitemOpen
  \bibfield  {author} {\bibinfo {author} {\bibfnamefont {F.}~\bibnamefont
  {Grosshans}}\ and\ \bibinfo {author} {\bibfnamefont {P.}~\bibnamefont
  {Grangier}},\ }\href {\doibase 10.1103/PhysRevA.64.010301} {\bibfield
  {journal} {\bibinfo  {journal} {Phys. Rev. A}\ }\textbf {\bibinfo {volume}
  {64}},\ \bibinfo {pages} {010301} (\bibinfo {year} {2001})}\BibitemShut
  {NoStop}%
\bibitem [{\citenamefont {Everitt}\ \emph {et~al.}(2004)\citenamefont
  {Everitt}, \citenamefont {Clark}, \citenamefont {Stiffell}, \citenamefont
  {Vourdas}, \citenamefont {Ralph}, \citenamefont {Prance},\ and\ \citenamefont
  {Prance}}]{ground_state}%
  \BibitemOpen
  \bibfield  {author} {\bibinfo {author} {\bibfnamefont {M.~J.}\ \bibnamefont
  {Everitt}}, \bibinfo {author} {\bibfnamefont {T.~D.}\ \bibnamefont {Clark}},
  \bibinfo {author} {\bibfnamefont {P.~B.}\ \bibnamefont {Stiffell}}, \bibinfo
  {author} {\bibfnamefont {A.}~\bibnamefont {Vourdas}}, \bibinfo {author}
  {\bibfnamefont {J.~F.}\ \bibnamefont {Ralph}}, \bibinfo {author}
  {\bibfnamefont {R.~J.}\ \bibnamefont {Prance}}, \ and\ \bibinfo {author}
  {\bibfnamefont {H.}~\bibnamefont {Prance}},\ }\href {\doibase
  10.1103/PhysRevA.69.043804} {\bibfield  {journal} {\bibinfo  {journal} {Phys.
  Rev. A}\ }\textbf {\bibinfo {volume} {69}},\ \bibinfo {pages} {043804}
  (\bibinfo {year} {2004})}\BibitemShut {NoStop}%
\bibitem [{\citenamefont {Banerji}(2001)}]{Banerji2001}%
  \BibitemOpen
  \bibfield  {author} {\bibinfo {author} {\bibfnamefont {J.}~\bibnamefont
  {Banerji}},\ }\href {\doibase 10.1007/s12043-001-0123-8} {\bibfield
  {journal} {\bibinfo  {journal} {Pramana}\ }\textbf {\bibinfo {volume} {56}},\
  \bibinfo {pages} {267} (\bibinfo {year} {2001})}\BibitemShut {NoStop}%
\bibitem [{\citenamefont {Hannay}\ and\ \citenamefont
  {Berry}(1980)}]{berry1980}%
  \BibitemOpen
  \bibfield  {author} {\bibinfo {author} {\bibfnamefont {J.}~\bibnamefont
  {Hannay}}\ and\ \bibinfo {author} {\bibfnamefont {M.}~\bibnamefont {Berry}},\
  }\href {\doibase https://doi.org/10.1016/0167-2789(80)90026-3} {\bibfield
  {journal} {\bibinfo  {journal} {Physica D: Nonlinear Phenomena}\ }\textbf
  {\bibinfo {volume} {1}},\ \bibinfo {pages} {267 } (\bibinfo {year}
  {1980})}\BibitemShut {NoStop}%
\bibitem [{\citenamefont {Clarke}\ and\ \citenamefont
  {Wilhelm}(2008)}]{superconducting_bits}%
  \BibitemOpen
  \bibfield  {author} {\bibinfo {author} {\bibfnamefont {J.}~\bibnamefont
  {Clarke}}\ and\ \bibinfo {author} {\bibfnamefont {F.~K.}\ \bibnamefont
  {Wilhelm}},\ }\href {http://dx.doi.org/10.1038/nature07128} {\bibfield
  {journal} {\bibinfo  {journal} {Nature}\ }\textbf {\bibinfo {volume} {453}},\
  \bibinfo {pages} {1031 EP } (\bibinfo {year} {2008})}\BibitemShut {NoStop}%
\bibitem [{\citenamefont {DiCarlo}\ \emph {et~al.}(2009)\citenamefont
  {DiCarlo}, \citenamefont {Chow}, \citenamefont {Gambetta}, \citenamefont
  {Bishop}, \citenamefont {Johnson}, \citenamefont {Schuster}, \citenamefont
  {Majer}, \citenamefont {Blais}, \citenamefont {Frunzio}, \citenamefont
  {Girvin},\ and\ \citenamefont {Schoelkopf}}]{squid_algo}%
  \BibitemOpen
  \bibfield  {author} {\bibinfo {author} {\bibfnamefont {L.}~\bibnamefont
  {DiCarlo}}, \bibinfo {author} {\bibfnamefont {J.~M.}\ \bibnamefont {Chow}},
  \bibinfo {author} {\bibfnamefont {J.~M.}\ \bibnamefont {Gambetta}}, \bibinfo
  {author} {\bibfnamefont {L.~S.}\ \bibnamefont {Bishop}}, \bibinfo {author}
  {\bibfnamefont {B.~R.}\ \bibnamefont {Johnson}}, \bibinfo {author}
  {\bibfnamefont {D.~I.}\ \bibnamefont {Schuster}}, \bibinfo {author}
  {\bibfnamefont {J.}~\bibnamefont {Majer}}, \bibinfo {author} {\bibfnamefont
  {A.}~\bibnamefont {Blais}}, \bibinfo {author} {\bibfnamefont
  {L.}~\bibnamefont {Frunzio}}, \bibinfo {author} {\bibfnamefont {S.~M.}\
  \bibnamefont {Girvin}}, \ and\ \bibinfo {author} {\bibfnamefont {R.~J.}\
  \bibnamefont {Schoelkopf}},\ }\href {http://dx.doi.org/10.1038/nature08121}
  {\bibfield  {journal} {\bibinfo  {journal} {Nature}\ }\textbf {\bibinfo
  {volume} {460}},\ \bibinfo {pages} {240 EP } (\bibinfo {year}
  {2009})}\BibitemShut {NoStop}%
\bibitem [{\citenamefont {Gambetta}\ \emph {et~al.}(2017)\citenamefont
  {Gambetta}, \citenamefont {Chow},\ and\ \citenamefont
  {Steffen}}]{squid_comp}%
  \BibitemOpen
  \bibfield  {author} {\bibinfo {author} {\bibfnamefont {J.~M.}\ \bibnamefont
  {Gambetta}}, \bibinfo {author} {\bibfnamefont {J.~M.}\ \bibnamefont {Chow}},
  \ and\ \bibinfo {author} {\bibfnamefont {M.}~\bibnamefont {Steffen}},\ }\href
  {\doibase 10.1038/s41534-016-0004-0} {\bibfield  {journal} {\bibinfo
  {journal} {npj Quantum Information}\ }\textbf {\bibinfo {volume} {3}},\
  \bibinfo {pages} {2} (\bibinfo {year} {2017})}\BibitemShut {NoStop}%
\end{thebibliography}
\end{document}